\documentclass[12pt,preprint]{aastex}

\shorttitle{Scaling of the damping time}
\shortauthors{Arregui, Ballester, \& Goossens}

\begin{document}

\title{On the scaling of the damping time for resonantly damped oscillations in coronal loops}

\author{I\~nigo Arregui\altaffilmark{1}, Jos\'e Luis Ballester\altaffilmark{1},  
Marcel Goossens\altaffilmark{2,1}}


\altaffiltext{1}{Departament de F\'{\i}sica, Universitat de les Illes Balears,
E-07122 Palma de Mallorca, Spain. Email: inigo.arregui@uib.es, dfsjlb0@uib.es}
\altaffiltext{2}{Centre Plasma Astrophysics, Katholieke Universiteit Leuven,
Leuven, B-3001, Belgium.  Email: marcel.goossens@wis.kuleuven.be}

\begin{abstract}
There is not as yet full agreement on the mechanism that causes the
rapid damping of the oscillations observed by TRACE in coronal
loops. It has been suggested that the variation of the observed
values of the damping time as function of the corresponding observed
values of the period contains information on the possible damping
mechanism. The aim of this Letter is to show that, for resonant absorption, this is definitely
not the case unless detailed a priori information on the individual
loops is available.
\end{abstract}


\section{Introduction}
Transverse oscillations in coronal loops have been detected in observations 
made with the EUV telescope on board of the Transition Region and Coronal
Explorer (TRACE) in 1999 by \citet{Aschwanden99} and
\citet{Nakariakov99}. Since then the detection of these
oscillations has been confirmed and in addition damped oscillations
have been observed in hot coronal loops by the SUMER instrument on
board SOHO (\citealt{Wang02,Kliem02}). The TRACE oscillations have periods of the order of
$\simeq 2 -10$ minutes and comparatively short damping times of the
order of $\simeq 3 -20$ minutes. There is general
consensus that these oscillations are fast standing kink mode
oscillations. However, there is still debate about the mechanism
that causes the observed fast damping. Different mechanisms have
been suggested: phase mixing (\citealt{OA02}), resonant
absorption (\citealt{HY88,GAA02,RR02}), lateral and foot-point wave 
leakage (\citealt{Smith97,depontieu01}), drag due
to the ambient plasma (\citealt{Chen07}). In order to discriminate
between different damping mechanisms \cite{OA02} suggested to
study how the observed damping times vary as a function of the
corresponding observed periods. In particular \cite{OA02}
claimed that the observed values are compatible with phase mixing if
the damping time increases with period $T$ as $T^{4/3}$ and with
resonant absorption if it increases as $T$. The aim of the present
Letter is to show that random observations of oscillation events in
coronal loops are very unlikely to produce any particular relation
between damping times and periods whatever the actual mechanism is
that causes the damping. In particular we focus on resonant
absorption and use synthetic data for periods and damping times to
show that various samples of pairs of periods and damping times can
lead to various and widely different scaling laws.

\section{Analytical theory}

The suggestion that the damping time increases linearly with period,
$\tau_{d}\sim T$, was triggered by  analytical asymptotic expressions for the damping
of the quasi-mode given by e.g. \cite{Goossens92},
\cite{RR02}, \cite{GAA02} (see reviews
\citealt{GAA06} and \citealt{Goossens08}). These asymptotic
expressions are derived in the approximation that the non-uniform
layer is thin.  This is the so-called thin boundary (TB) approximation. 
Jump conditions are used to connect the solution over the ideal
singularity and to avoid solving the non-ideal MHD wave equations 
(see e.g. \citealt{sakurai91,goossens95}).
The asymptotic expression for the damping time in the TB approximation is

\begin{equation}
\frac{\displaystyle \tau_{d}}{\displaystyle T} = F \;\; \frac {\displaystyle R}
{\displaystyle l}\;\; \frac{\displaystyle \zeta + 1}{\displaystyle
\zeta - 1}, \label{DecayTimeRR2}
\end{equation}

\noindent
where $\zeta=\rho_i / \rho_e$ is the density contrast,  $\rho_i$ and
$\rho_e$ the constant internal and  external densities, $l$ the thickness of the 
non-uniform layer defined in [$R-l/2$, $R+l/2$], and $R$ the mean radius.  
$F$ is a parameter that depends on the variation of density in the transitional layer. 
For a linear variation $F=4/\pi^2$  (\citealt{HY88,Goossens92}), for a sinusoidal 
variation $F=2/\pi$  (\citealt{RR02}). In what follows we shall use the version with $F=2/\pi$.

Equation~(\ref{DecayTimeRR2}) might be considered to be an
invitation for predicting a linear variation of damping time with
period. However, the crucial point for this prediction to make sense
is that the remaining quantities in equation~(\ref{DecayTimeRR2}) are
constants; i.e. all the oscillations events under considerations
have to occur in loops with the same density contrast $\zeta$ and
the same radial inhomogeneity $l/R$. When applied to a collection of
oscillating loops on which no prior information on their structure
is available equation~(\ref{DecayTimeRR2}) does not allow any prediction. A
naive use of  equation~(\ref{DecayTimeRR2}) indicates that for a given period
$T$  the damping times can differ by f.e. a factor $50/3$ for the
combinations $\zeta =1.5, l/R = 0.1$ and  $\zeta =5, l/R = 0.5$.

This is indeed a naive use of equation~(\ref{DecayTimeRR2}) since $T$ is a
function of $\zeta$ also. In order to make that clear we  use results
from the thin tube (TT) or long wave-length approximation for the
period $T$. Our starting point is the well-known expression for the
square of the frequency of the kink mode in a cylinder with
a uniform and straight magnetic field along the $z$-axis (see e.g. \citealt{ER83})

\begin{equation}
\omega^2 = \omega_{k}^2 =  \frac {\displaystyle \rho_i
\omega_{A,i}^2 + \rho_e  \omega_{A,e}^2} {\displaystyle \rho_i +
\rho_e}, \label{TrueDisc1}
\end{equation}

\noindent
where $\omega_{A}=k_z v_A$ is the local  Alfv\'{e}n frequency and 
$v_A=B/\sqrt{\mu \rho}$  the local  Alfv\'{e}n velocity (the subscripts 
``i'' and ``e'' refer to internal and external, respectively).
Now we note that for the fundamental mode $k_z = \pi/L$, with $L$ the length of the loop, and
we convert frequencies to periods and rewrite equation~(\ref{TrueDisc1}) as

\begin{equation}
\frac{\displaystyle T}{\displaystyle \tau_{A,i}} =
\sqrt{2}  \; A(\zeta),
\label{PeriodNorm1}
\end{equation}

\noindent
where $\tau_{A,i}=L/v_{A,i}$ is the internal Alfv\'{e}n travel time
and the function $A(\zeta)$ is defined as

\begin{equation}
A(\zeta) = \left ( \frac{\displaystyle \zeta +1}{\displaystyle
\zeta} \right )^{1/2}. \label{A}
\end{equation}

It is convenient to adopt a reference loop with magnetic field strength $B_0$, 
length $L_0$, and internal density $\rho_{i0}$. This reference loop allows us to consider loops of
different dimensions  and to introduce the normalized period $T^{\star}$ defined as

\begin{equation}
T^{\star} = \frac{\displaystyle T}{\displaystyle
(\sqrt{2})\tau_{A,i0}}. \label{PeriodNorm2}
\end{equation}

\noindent
$T^{\star}$ can be interpreted as the period measured in the unit
$\tau_{A,i0} \sqrt{2}$.   With this normalized period
equation~(\ref{PeriodNorm1}) takes the simple form

\begin{equation}
T^{\star} = a \; A(\zeta),  \label{PeriodNorm3}
\end{equation}

\noindent
with $a=\tau_{A,i}/\tau_{A,i0}$. From equation~(\ref{PeriodNorm3}) we learn that, 
for a given value of $a$, $T^{\star}$ varies continuously from $\sqrt{2} a$ 
to $a$ when $\zeta$ is allowed to vary from 1 (no loop) to $\infty$ (outside
vacuum).

Let us now turn back to equation~(\ref{DecayTimeRR2}) and rewrite it as

\begin{equation}
\frac{\displaystyle \tau_{d}}{\displaystyle \tau_{A,i}} = \frac
{\displaystyle 2 \sqrt{2}} {\displaystyle \pi}   \; B(\zeta) \;
\frac{\displaystyle 1}{\displaystyle l/R}, \label{DecayTNorm1}
\end{equation}

\noindent
with the function $B(\zeta)$ defined as

\begin{equation}
B(\zeta) = \left ( \frac{\displaystyle \zeta +1}{\displaystyle
\zeta} \right )^{1/2} \frac{\displaystyle \zeta +1}{\displaystyle
\zeta - 1} = A(\zeta) \frac{\displaystyle \zeta +1}{\displaystyle
\zeta - 1}. \label{B}
\end{equation}

\noindent
In what follows we shall refer to the combination of the TB
approximation to compute the damping time and the TT approximation
for the period as the TTTB approximation.

In the same way as  for the period it is convenient to introduce the
normalized decay time $\tau_{d}^{\star}$ as

\begin{equation}
\tau_{d}^{\star} = \frac{\displaystyle \tau_{d} }
{\displaystyle (2 \sqrt{2}/ \pi) \tau_{A,i0}}.
\label{DecayTNorm2}
\end{equation}

\noindent
$\tau_{d}^{\star}$ can be interpreted as the decay time measured in
the unit $\tau_{A,i0} \; (2 \sqrt{2}/ \pi) $.  With
the use of this normalized decay time we can write
equation~(\ref{DecayTNorm1}) as

\begin{equation}
\tau_{d}^{\star} = a \; B(\zeta) \; \frac{\displaystyle
1}{\displaystyle l/R}, \label{DecayTNorm3}
\end{equation}

\noindent
which tells us that, for a given value of $a$,
$\tau_{d}^{\star}$ depends on $\zeta$ and $l/R$. The dependency on
$l/R$ is straightforward. The dependency on $\zeta$ is slightly more
complicated. Basically, $\tau_{d}^{\star}$ is a decreasing function
of $\zeta$ varying from $+ \infty$ for $\lim \zeta \rightarrow 1$ to
$\frac{a}{l/R}$ for $\lim \zeta \rightarrow + \infty$. In summary,
$\tau_{d}^{\star}$ is a decreasing function of both $l/R$ and
$\zeta$.

\begin{figure}
\plotone{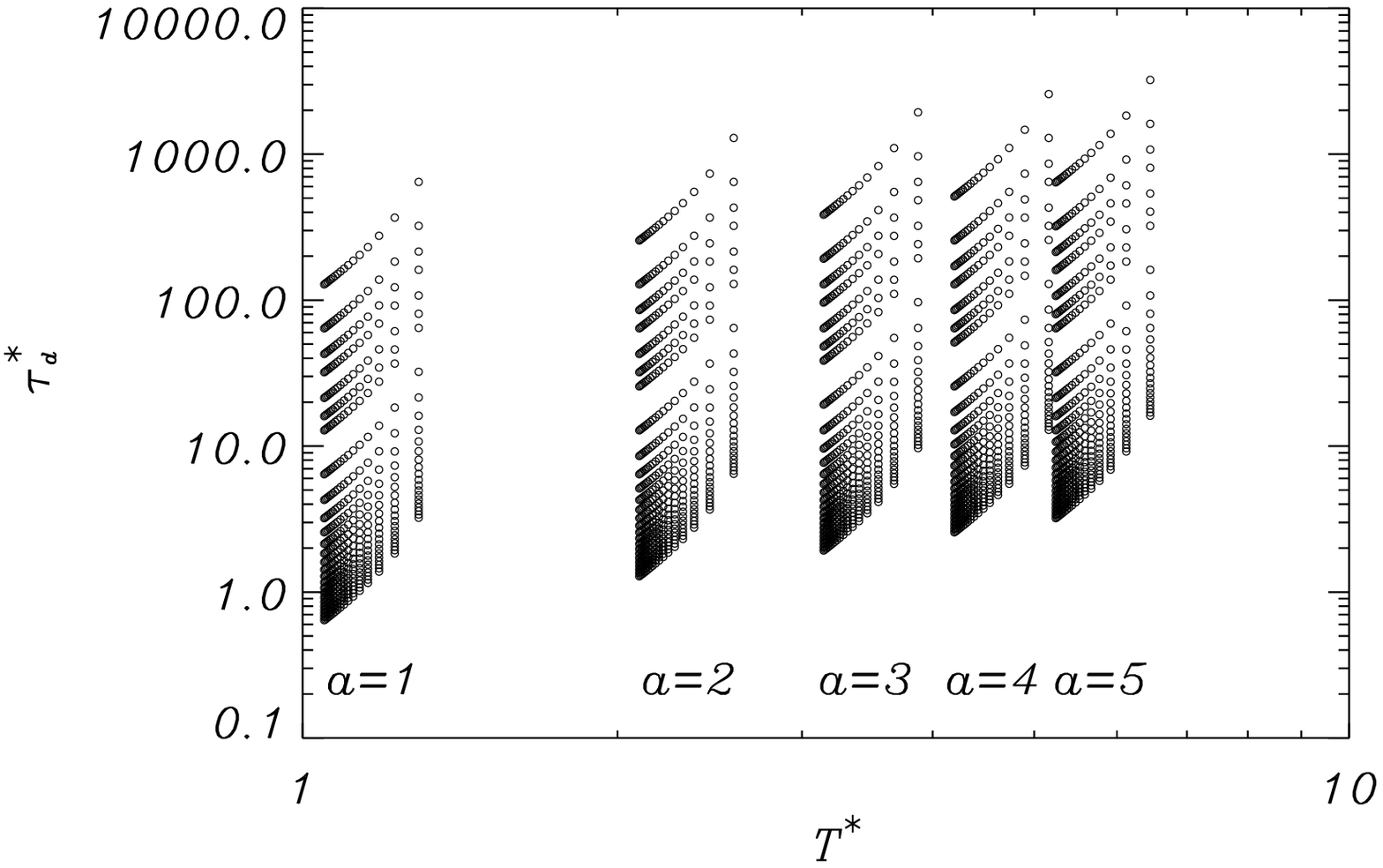}
\caption{Normalized damping time as a function of the normalized period 
for discrete values of $\zeta\in[1.5,10]$ and $l/R\in[0.01,2]$, and five values of $a$.}
\label{fig1}
\end{figure}

We now combine equations~(\ref{PeriodNorm3}) and  (\ref{DecayTNorm3}). On
Figure~\ref{fig1} we have plotted $\tau_{d}^{\star}$  versus
$T^{\star}$  for values of $\zeta$ from 1.5 up to 10 and  of $l/R$
from 0.01 to 2. For every value of $a$
the pairs $( T^{\star},  \tau_{d}^{\star})$ define a vertical strip
in the $(T^{\star},  \tau_{d}^{\star})$-plane with a maximum horizontal
extent of $(\sqrt{2} - 1) a$. As a matter of fact $a$ can take on
any positive real value, but for clarity we have limited the values
 to $a=1,2,3,4,5$. The inclusion of intermediate
values of $a$, such as $a = 4/3, 5/3, 7/3$, produces  vertical strips
overlapping those shown on Figure~\ref{fig1}. The iso-lines corresponding to a
constant value of $\zeta$ are vertical lines since the period is independent of the inhomogeneity
length scale, in the thin tube
approximation. The iso-lines corresponding to a constant value of $l/R$ are
not straight and defined by the equation

\begin{equation}
\tau_{d}^{\star} = \frac{\displaystyle (T^{\star})^3}{ 2 a^2 -
(T^{\star})^2 } \; \frac{\displaystyle 1}{\displaystyle l/R} \;\;\;
\mbox{for} \;\;T^{\star} \in [a, a \sqrt{2}[ .\label{Isolinel}
\end{equation}

\noindent
It is clear from Figure~\ref{fig1} that the model of resonant absorption in
its most simple mathematical formulation using the TTTB
approximation produces a wide variety of combinations of $(
T^{\star}, \tau_{d}^{\star}).$  Now let us see what happens when a
collection of pairs $( T^{\star}, \tau_{d}^{\star})$ is drawn from
this reservoir.

\section{Scaling laws: how many do we want?}

\begin{figure}
\plotone{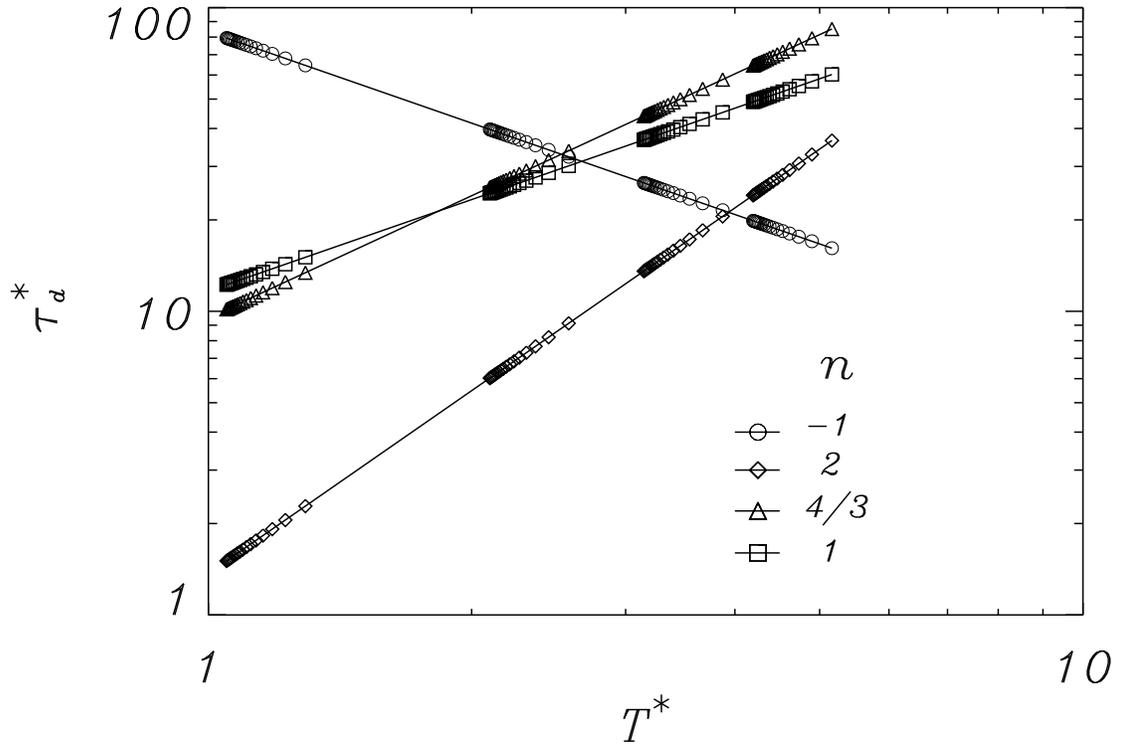}
\caption{Four synthetic scaling laws corresponding to $n = -1, 4/3, 1,2$, 
and values of $a$, from left to right, $a=1, 2, 3, 4$.}
\label{fig2}
\end{figure}

The aim of this section is to show that different collections of
data can be produced by the simple TTTB mathematical model of
resonant absorption with each of them leading to scaling laws, 
$\tau^{\star}_d/(T^{\star})^n=C$ (with $C$ a function of the equilibrium parameters), with
different indices, $n$. 
We  use equations~(\ref{PeriodNorm3}) and (\ref{DecayTNorm3}),  together with the particular 
relation between $\tau^{\star}_d$ and $T^{\star}$ for each index $n$, and derive those collections 
of data. As a first example we have plotted on Figure~\ref{fig2} a
collection of $( T^{\star}, \tau_{d}^{\star})$ combinations drawn
from the big reservoir depicted on Figure~\ref{fig1}. These combinations
define a  scaling law with index $n=-1$. It is clear that we have
engineered the data presented on Figure~\ref{fig2} so as to fit the scaling
law with such index . Actually, the engineering is relatively
straightforward. We require $T^{\star}\tau^{\star}_d=C_s$. The left-hand side is 
the value of the function $C$ for a starting configuration 
with $a=a_S$, $\zeta=\zeta_S$, and  $l/R=(l/R)_S$. We fix a loop to start with by prescribing the
values  $a_S= 1$, $\zeta_S= 1.5$, and $(l/R)_S=0.1$ 
and compute values of $a$, $\zeta$, and $l/R$ that satisfy the
equation
\begin{equation}
\frac{\displaystyle l}{\displaystyle R} = \left (\frac{\displaystyle
l}{\displaystyle R}\right )_S \; \left (\frac{\displaystyle
a}{\displaystyle a_S}\right )^2 \; \frac{\displaystyle
f(\zeta)}{\displaystyle f(\zeta_S)}; \mbox{\hspace{0.5cm}}f(\zeta) = A(\zeta) \; B(\zeta).  \label{Scaling-1}
\end{equation}

Likewise on Figure~\ref{fig2} we have also 
plotted a second collection of data that now define a
scaling law with index $n=2$. To obtain these data, we start from a loop with
prescribed values $a_S= 1$, $\zeta_S= 5$, and $(l/R)_S= 1$ and compute values of
$a$, $\zeta$, and $l/R$ that satisfy the equation
\begin{equation}
\frac{\displaystyle l}{\displaystyle R} = \left (\frac{\displaystyle
l}{\displaystyle R}\right )_S \; \frac{\displaystyle
a_S}{\displaystyle a} \; \frac{\displaystyle g(\zeta)}{\displaystyle
g(\zeta_S)};  \mbox{\hspace{0.5cm}} g(\zeta) = \frac{\displaystyle B(\zeta)} {\displaystyle \left (
A(\zeta) \right )^2}. \label{Scaling+2}
\end{equation}

A third collection of data, that now define a
scaling law with index $n=4/3$ is also shown in Figure~\ref{fig2}. According to
\cite{OA02} this value of the index singles out phase mixing as
damping mechanism. Here it is obtained for a collection of data that
are produced by the theoretical prediction for resonant absorption.
We start from a loop with prescribed values  $a_S= 1$,
$\zeta_S= 3$, and $(l/R)_S= 0.5$ and compute values of $a$, $\zeta$, and $l/R$ that
satisfy the equation

\begin{equation}
\frac{\displaystyle l}{\displaystyle R} = \left (\frac{\displaystyle
l}{\displaystyle R}\right )_S \; \left (\frac{\displaystyle
a_S}{\displaystyle a} \right )^{1/3} \; \frac{\displaystyle
h(\zeta)}{\displaystyle h(\zeta_S)};  \mbox{\hspace{0.25cm}} h(\zeta) = \frac{\displaystyle B(\zeta)}
{\displaystyle \left ( A(\zeta) \right )^{4/3}}.\label{Scaling4/3}
\end{equation}

Finally, on Figure~\ref{fig2} we have also plotted a collection of data that
define a scaling law with the canonical value n=1 for the index.  We start from a loop
with prescribed values $a_S= 1$, $\zeta_S= 2.5$, and $(l/R)_S= 0.2$ and compute values
of $a$, $\zeta$, and $l/R$ that satisfy the equation

\begin{equation}
\frac{\displaystyle l}{\displaystyle R} = \left( \frac{\displaystyle
l}{\displaystyle R}\right)_S \;\frac{\displaystyle c(\zeta)}
{\displaystyle c(\zeta_S)};   \mbox{\hspace{0.5cm}}c(\zeta) = \frac{\displaystyle  B(\zeta)} {\displaystyle A(\zeta)} =
\frac{\displaystyle \zeta +1}{\displaystyle \zeta - 1}.\label{Scaling+1}
\end{equation}

\noindent
Note that $a$ is absent from (\ref{Scaling+1}) meaning that a
solution of (\ref{Scaling+1}) can combined with any value of $a>0$.
An obvious solution to (\ref{Scaling+1}) is
\begin{equation}
\zeta = \zeta_S, \;\;\frac{\displaystyle l}{\displaystyle R} =
\left( \frac{\displaystyle l}{\displaystyle R}\right)_S,\;\; a >0.
\label{Scaling+1SS}
\end{equation}

\noindent
Equation~(\ref{Scaling+1SS}) defines a straight line in the $( T^{\star},  \tau_{d}^{\star}) $-plane. 
Note that equation~(\ref{Scaling+1SS}) is not the only solution in the $( T^{\star}, \tau_{d}^{\star}) 
$-plane. Any combination $(T^{\star}, \;\tau_{d}^{\star})$ that satisfies 
equation~(\ref{Scaling+1}) combined with $a \in ]0, \; \infty[$ produces the same 
straight line in the  $( T^{\star},  \tau_{d}^{\star}) $
plane as solution (\ref{Scaling+1SS}). Since the function $c(\zeta)$
is a decreasing function of its argument, a value $\zeta > \zeta_S$
is combined with a value $l/R < (l/R)_S$ and conversely a value
$\zeta < \zeta_S$ is combined with a value $l/R > (l/R)_S$.

It is clear from Figure~\ref{fig2}  that almost any scaling law can be
obtained from data produced by the simple TTTB mathematical model of
resonant absorption. All the periods and damping times plotted in Figure~\ref{fig2} correspond to 
combinations of  $\zeta$ and $l/R$, obtained from equations~(\ref{Scaling-1}), (\ref{Scaling+2}), 
(\ref{Scaling4/3}), and (\ref{Scaling+1}),  that are reasonable values  of these quantities in the ranges $\zeta\in[1.5,10]$ and $l/R\in[0.016,1.59]$. The discrete sets of values in Figure~\ref{fig2} arise due to the considered discrete values of $a$, but 
loops with different internal travel times, with respect to the reference loop, would give the full set of values for $\tau^{\star}_d$ and $T^{\star}$ along the solid lines depicted in Figure~\ref{fig2}.

\section{Beyond the TTTB approximation}

\begin{figure}
\plotone{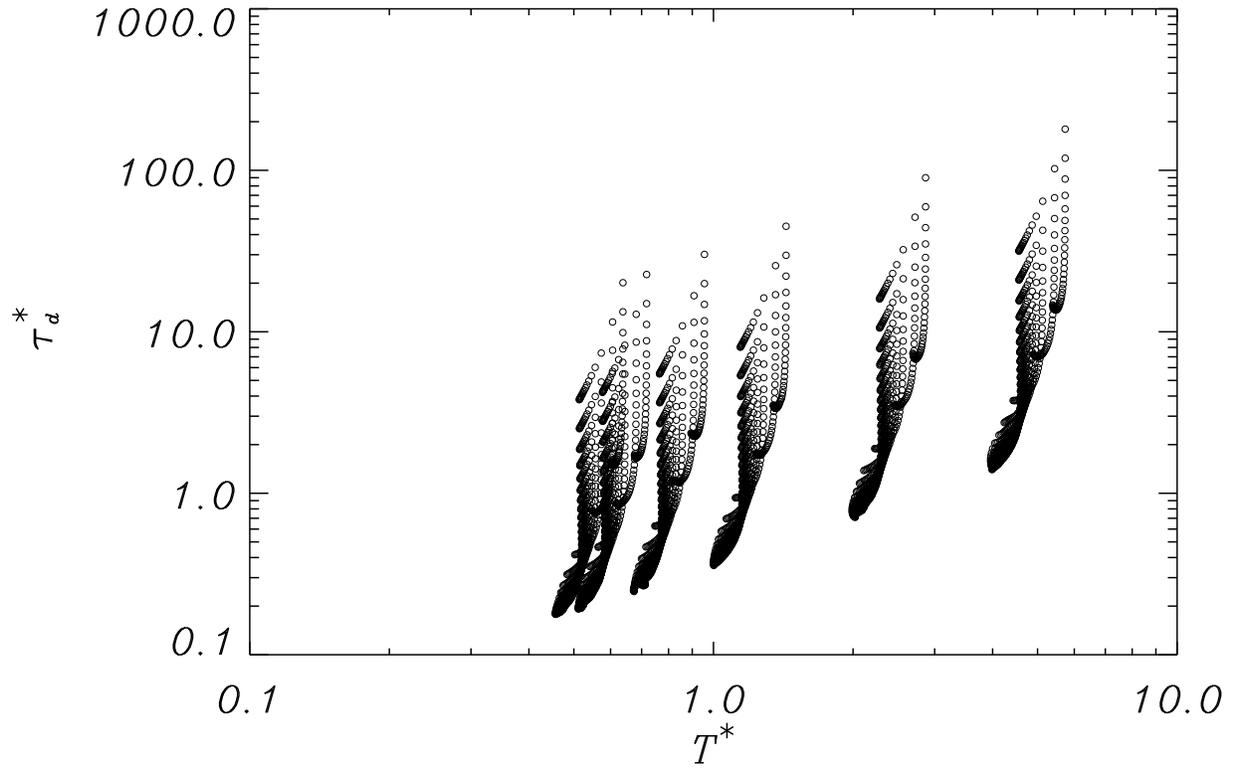}
\caption{Normalized damping time as a function of the normalized period for fully non-uniform 
loop models with $\zeta\in[1.5,20]$ and $l/R\in[0.1,2]$.
The six vertical strips correspond to six values of $k_z=\pi/L$.}
\label{fig3}
\end{figure}

The TTTB approximation  as a
mathematical model for resonant absorption has its clear
limitations. First, the values of the period defined by
equation~(\ref{PeriodNorm3}) are independent of the radius $R$ and the 
inhomogeneity length scale $l/R$. In reality, the period is a function of  the
density contrast $\zeta$ , $l/R$, and $R$. As a consequence the iso-lines
corresponding to a constant value of $\zeta$ in the  $( T^{\star},
\tau_{d}^{\star})$-plane  are no longer straight vertical lines.  For
a given value of $a$ this dependency of the period on $\zeta$
produces additional spread on the original vertical strips. Second,
the TTTB approximation  is an accurate  approximation as long as the
damping time is sufficiently longer than the period, since this is
an inherent assumption for carrying out the asymptotic analysis
leading to equation~(\ref{DecayTimeRR2}). We can expect this equation 
to be inaccurate for large values of the density contrast $\zeta$ 
and large values of the inhomogeneity length scale $l/R$. 
Figure~\ref{fig3} is the twin version of Figure~\ref{fig1}
with the values of $T^{\star}$ and $\tau_{d}^{\star}$ now computed
for fully non-uniform 1-D loops  (\citealt{tom04b}). Figure~\ref{fig3} is very similar in
appearance to Figure~\ref{fig1}. The differences occur where they expected to
occur. The vertical strips of Figure~\ref{fig1} are now replaced with strips
slightly oblique to the vertical axis and the largest differences
appear at the short values of the damping times corresponding to the
large values of the density contrast $\zeta$ and large values of the
inhomogeneity length scale $l/R$. The basic message from both figures is the same.
The model of resonant absorption produces a wide variety of combinations in 
the $( T^{\star}, \tau_{d}^{\star})$-plane. If we draw a curve defined by a given 
relation between $T^{\star}$ and $\tau_{d}^{\star}$, for instance $ \tau_{d}^{\star} /
(T^{\star})^2 = C$,  in this plane we can graphically determine as
many points on this curve as we want. If we select the data
determined in this way the result is a scaling law with index $n=2$ 
as the one in Figure~\ref{fig2}. The only difference is that now
the procedure is numerical all the way.

\section{Conclusion}In this Letter we have explained that scaling
laws for a group of observations of oscillating loops to
discriminate between different damping mechanisms are of not much
use if there is no a priori information on the properties of the
loops. The analytical expressions obtained by the TTTB
approximation enable us to show in a straightforward and easy to
follow manner that in the framework of resonant absorption
collections of synthetic data can be produced that follow almost any
scaling law. Then, we have backed up our findings by numerical
eigenvalue computations that do not suffer from the TTTB limitations but are on their own rather
less instructive.

It might be argued that nature is not as
cunning as the authors of this paper and does not attempt to
engineer data according to prescribed rules as those defined in these 
equations. On the other hand, there is no reason why nature
would want to produce coronal loops that all have the same values of
$\zeta$ and $l/R$. So, unless there is accurate a priori information
on the coronal loops available, the use of scaling laws to
discriminate between different damping mechanisms is questionable, to say the least.

\acknowledgments
It is pleasure for MG to acknowledge the warm hospitality of JLB, the friendly atmosphere
of the Solar Physics Group at the UIB, and the visiting position from the UIB. 
MG also acknowledges the FWO-Vlaanderen for awarding him a sabbatical leave.
IA and JLB acknowledge the funding provided under projects AYA2006-07637 (Spanish MEC)
and PRIB-2004-10145 and PCTIB2005GC3-03 (Government of the Balearic Islands). The 
authors are grateful to  R. Oliver, J. Terradas, and T. Van Doorsselaere
for  comments and suggestions.

\clearpage

\clearpage

\clearpage

\clearpage

\begin{thebibliography}{18}
\expandafter\ifx\csname natexlab\endcsname\relax\def\natexlab#1{#1}\fi

\bibitem[{{Aschwanden} {et~al.}(1999){Aschwanden}, {Fletcher}, {Schrijver}, \&
  {Alexander}}]{Aschwanden99}
{Aschwanden}, M.~J., {Fletcher}, L., {Schrijver}, C.~J., \& {Alexander}, D.
  1999, \apj, 520, 880

\bibitem[{{Chen} \& {Schuck}(2007)}]{Chen07}
{Chen}, J. \& {Schuck}, P.~W. 2007, \solphys, 246, 145

\bibitem[{{De Pontieu} {et~al.}(2001){De Pontieu}, {Martens}, \&
  {Hudson}}]{depontieu01}
{De Pontieu}, B., {Martens}, P.~C.~H., \& {Hudson}, H.~S. 2001, \apj, 558, 859

\bibitem[{{Edwin} \& {Roberts}(1983)}]{ER83}
{Edwin}, P.~M. \& {Roberts}, B. 1983, \solphys, 88, 179

\bibitem[{{Goossens}(2008)}]{Goossens08}
{Goossens}, M. 2008, in Proceedings IAU Symposium No. 247, ed. R.~{Erdelyi} \&
  C.~{Mendoza}

\bibitem[{{Goossens} {et~al.}(2006){Goossens}, {Andries}, \& {Arregui}}]{GAA06}
{Goossens}, M., {Andries}, J., \& {Arregui}, I. 2006, Royal Society of London
  Philosophical Transactions Series A, 364, 433

\bibitem[{{Goossens} {et~al.}(2002){Goossens}, {Andries}, \&
  {Aschwanden}}]{GAA02}
{Goossens}, M., {Andries}, J., \& {Aschwanden}, M.~J. 2002, \aap, 394, L39

\bibitem[{{Goossens} {et~al.}(1992){Goossens}, {Hollweg}, \&
  {Sakurai}}]{Goossens92}
{Goossens}, M., {Hollweg}, J.~V., \& {Sakurai}, T. 1992, \solphys, 138, 233

\bibitem[{{Goossens} {et~al.}(1995){Goossens}, {Ruderman}, \&
  {Hollweg}}]{goossens95}
{Goossens}, M., {Ruderman}, M.~S., \& {Hollweg}, J.~V. 1995, \solphys, 157, 75

\bibitem[{{Hollweg} \& {Yang}(1988)}]{HY88}
{Hollweg}, J.~V. \& {Yang}, G. 1988, \jgr, 93, 5423

\bibitem[{{Kliem} {et~al.}(2002){Kliem}, {Dammasch}, {Curdt}, \&
  {Wilhelm}}]{Kliem02}
{Kliem}, B., {Dammasch}, I.~E., {Curdt}, W., \& {Wilhelm}, K. 2002, \apjl, 568,
  L61

\bibitem[{{Nakariakov} {et~al.}(1999){Nakariakov}, {Ofman}, {DeLuca},
  {Roberts}, \& {Davila}}]{Nakariakov99}
{Nakariakov}, V.~M., {Ofman}, L., {DeLuca}, E.~E., {Roberts}, B., \& {Davila},
  J.~M. 1999, Science, 285, 862

\bibitem[{{Ofman} \& {Aschwanden}(2002)}]{OA02}
{Ofman}, L. \& {Aschwanden}, M.~J. 2002, \apjl, 576, L153

\bibitem[{{Ruderman} \& {Roberts}(2002)}]{RR02}
{Ruderman}, M.~S. \& {Roberts}, B. 2002, \apj, 577, 475

\bibitem[{{Sakurai} {et~al.}(1991){Sakurai}, {Goossens}, \&
  {Hollweg}}]{sakurai91}
{Sakurai}, T., {Goossens}, M., \& {Hollweg}, J.~V. 1991, \solphys, 133, 227

\bibitem[{{Smith} {et~al.}(1997){Smith}, {Roberts}, \& {Oliver}}]{Smith97}
{Smith}, J.~M., {Roberts}, B., \& {Oliver}, R. 1997, \aap, 317, 752

\bibitem[{{Van Doorsselaere} {et~al.}(2004){Van Doorsselaere}, {Andries},
  {Poedts}, \& {Goossens}}]{tom04b}
{Van Doorsselaere}, T., {Andries}, J., {Poedts}, S., \& {Goossens}, M. 2004,
  \apj, 606, 1223

\bibitem[{{Wang} {et~al.}(2002){Wang}, {Solanki}, {Curdt}, {Innes}, \&
  {Dammasch}}]{Wang02}
{Wang}, T., {Solanki}, S.~K., {Curdt}, W., {Innes}, D.~E., \& {Dammasch}, I.~E.
  2002, \apjl, 574, L101

\end{thebibliography}
\end{document}